\documentstyle[twoside,fleqn,npb,epsfig]{article}

\newcommand{\beq}{\begin{eqnarray}}
\newcommand{\eeq}{\end{eqnarray}}

\newcommand{\AmS}{{\protect\the\textfont2
  A\kern-.1667em\lower.5ex\hbox{M}\kern-.125emS}}

\title{Twist-3 effects in polarized Drell-Yan and semi-inclusive DIS}

\author{Yuji Koike\address{Department of Physics, Niigata University,
        Ikarashi, Niigata 950-2181, Japan}}

\begin{document}

\begin{abstract}

Two twist-3 processes are discussed.  
We first present a simple estimate of the longitudinal-transverse spin
asymmetry in the polarized Drell-Yan process.  
We next derive a cross section for the
semi-inclusive production of a polarized spin-1/2 baryon
in the DIS of an unpolarized electron off the polarized nucleon.

\end{abstract}

\maketitle

Spin dependent processes in high energy scatterings
provide us with a possibility to 
extract twist-3 distribution and fragmentation
functions which represent
quark-gluon correlations inside the hadrons.
A well known example is the nucleon's transverse spin-structure
function $g_2(x,Q^2)$.
In this report, we discuss two other processes
in which twist-3 distribution functions reveal themselves as
a leading contribution with respect to the inverse power of 
the hard momentum $Q$.  The first one is the
longitudinal ($L$) and the transverse ($T$) spin asymmetry $A_{LT}$ in the 
nucleon-nucleon polarized Drell-Yan process.  We present a simple estimate
for $A_{LT}$ in comparison with $A_{LL}$ and $A_{TT}$.\cite{KKN}  
The second one is the semi-inclusive production of a polarized spin-1/2
baryon in the deep inelastic scattering from a polarized nucleon.
A formula for the cross section is presented.

In LO QCD, the double spin asymmetries in the nucleon-nucleon
polarized Drell-Yan process
are given by\,\cite{JJ}
\beq
A_{LL}N \hskip-0.25cm
       &=&\hskip-0.25cm  \Sigma_{a} e_{a}^2
                   g_{1}^a(x_1,Q^2)g_{1}^{\bar{a}}(x_2,Q^2),
\label{ALL}\\[5pt]
A_{TT}N \hskip-0.25cm
&=& \hskip-0.25cm a_{TT} \Sigma_{a} e_{a}^2
                   h_{1}^a(x_1,Q^2)h_{1}^{\bar{a}}(x_2,Q^2),
\label{ATT}\\[5pt]
A_{LT}N \hskip-0.25cm
       &=& \hskip-0.25cm a_{LT} 
\Sigma_{a} e_{a}^2
        \left [g_{1}^a(x_1,Q^2)x_2g_{T}^{\bar{a}}(x_2,Q^2)\right.\nonumber\\
& &\left.    + x_1h_{L}^a(x_1,Q^2)h_{1}^{\bar{a}}(x_2,Q^2) \right],
\label{ALT}
\end{eqnarray}
where $N=\Sigma_{a} e_{a}^2 f_{1}^a(x_1,Q^2)f_{1}^{\bar{a}}(x_2,Q^2)$,
$Q^2$ is the virtuality of the photon, 
$e_a$ represent the electric charge of the quark-flavor
$a$ and the summation is over all quark and anti-quark flavors: 
$a=u,d,s,\bar{u},\bar{d},\bar{s}$, ignoring heavy quark 
contents ($c,b,\cdots$) 
in the nucleon.
The variables $x_1$ and $x_2$ refer to
the momentum fractions of the partons coming from
the two nucleons ``1'' and ``2'', respectively.
In (\ref{ATT}) and (\ref{ALT}), $a_{TT}$ and $a_{LT}$ represent
the asymmetries in the parton level defined as
$a_{TT}  =  {\rm sin}^2\theta\, 
{\rm cos}2\phi/(1+{\rm cos}^2\theta)$ and 
$a_{LT}  =  (M/Q) (2\,{\rm sin}2\theta\, 
{\rm cos}\phi)/(1+{\rm cos}^2\theta)$,
where $\theta$ and $\phi$ are, respectively, the
polar and azimuthal angle of the lepton momentum
with respect to the beam direction and 
the transverse spin.
We note that $A_{LL}$ and $A_{TT}$ receive contribution only from
the twist-2 distributions, while $A_{LT}$ is proportional to the twist-3 
distributions
and hence $a_{LT}$ is suppressed by a factor $1/Q$.

The twist-3 distributions $g_T$ and $h_L$ can be decomposed into
the twist-2 contribution  
and the ``purely twist-3'' contribution (at $m_q=0$):
\beq
g_{T}(x,\mu^2)\hskip-0.25cm &=&\hskip-0.25cm 
\int_{x}^{1} dy\frac{g_{1}(y,\mu^2)}{y}  + 
\widetilde{g}_{T}(x,\mu^2),\nonumber\\
h_{L}(x,\mu^2)\hskip-0.25cm &=&\hskip-0.25cm 
2x\int_{x}^{1}dy \frac{h_{1}(y,\mu^2)}{y^2}  
+ \widetilde{h}_{L}(x,\mu^2).
\label{gTWW}
\eeq
The purely twist-3 pieces
$\widetilde{g}_T$ and
$\widetilde{h}_L$ can be written as 
quark-gluon-quark correlators on the lightcone.
In the following we call the 
first terms in (\ref{gTWW}) 
$g_T^{WW}(x,\mu^2)$ and $h_L^{WW}(x,\mu^2)$ (Wandzura-Wilczek parts).

For the present estimate of $A_{LT}$,
we have used 
the LO parametrization by 
Gl\"uck-Reya-Vogt for $f_1$\,\cite{GRV}, 
the
LO parametrization (standard scenario)
by Gl\"uck-Reya-Stratmann-Vogelsang (GRSV)
and the LO model-A by Gehrmann and Stirling for $g_1$\,\cite{GRSV}.
For $h_1$, $g_T$ and $h_L$ no experimental
data is available up to now and we have to rely on some theoretical
postulates. 
Here 
we assume $h_1(x,\mu^2)=g_1(x,\mu^2)$ at a low energy scale
($\mu^2=0.23$ GeV$^2$) 
as has been suggested by a low energy nucleon model\,\cite{JJ}.
These assumptions also fix $g_T^{WW}$ and $h_L^{WW}$.
For the purely twist-3 parts $\widetilde{g}_T$ and
$\widetilde{h}_L$ we employ the bag model results at a low energy 
scale, assuming the bag scale is $\mu_{bag}^2=0.081$ and $0.25$ GeV$^2$.  
%In particular, we set the strangeness contributions
%to the purely twist-3 contributions equal to zero.
For the $Q^2$ evolution of $\widetilde{g}_T(x,Q^2)$ and 
$\widetilde{h}_L(x,Q^2)$,
we apply the large-$N_c$ evolution\cite{ABH}: In the $N_c\to \infty$ limit,
their evolution equation is reduced to a simple DGLAP form similarly to the
twist-2 distibution and a 
correction due to the finite 
value of $N_c$ is of $O(1/N_c^2)\sim 10$ \% level.

\begin{figure}[htb]
\vspace{-9pt}
\begin{center}
\mbox{~\epsfig{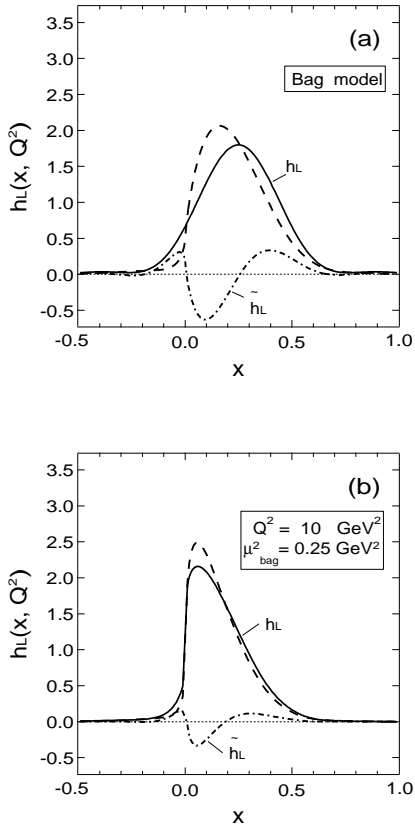}}
\caption{Bag model calculation of $h_L(x)$. 
The full, dashed, and dash-dot
lines, respectively, denote $h_L$, 
$h_L^{WW}$, and $\widetilde{h}_L$.}
\label{fig1}
\end{center}
\vspace{-19pt}
\end{figure}

To see the characteristics of the $Q^2$ evolution of the
twist-3 distributions, we show in  
Fig. 1 the bag model result for $h_L$ and
the one at $Q^2=10$ GeV $^2$ obtained by 
assuming that the bag scale is $\mu_{bag}^2 = 0.25$ $GeV^2$\,\cite{KK}.
One sees clearly that even though $\widetilde{h}_L(x,Q^2)$
is of the same order as $h^{WW}_L(x,Q^2)$ at a low energy scale,
$h_L(x,Q^2)$ is dominated by $h^{WW}_L(x,Q^2)$ at a high energy scale.
This feature is common to $g_T(x,Q^2)$ and
has a following general ground not peculiar to the bag model
calculation: 
\begin{enumerate}
\item[$\bullet$]
The anomalous dimensions for $\widetilde{h}_L(x,Q^2)$ and 
$\widetilde{g}_T(x,Q^2)$ 
are larger than those
for $h_1(x,Q^2)$ and $g_1(x,Q^2)$, hence faster $Q^2$ evolution for the former.
\item[$\bullet$]
$\widetilde{h}_L(x,Q^2)$ has a node.  This is due to
the sum rule
$\int_{-1}^1\,dx\,\widetilde{h}_L(x,Q^2)=0$ and 
$\int_{-1}^1\,dx\,\widetilde{g}_T(x,Q^2)=0$, 
which is known as 
the Burkhardt and Cottingham sum rule for $g_T(x,Q^2)$.
\end{enumerate}
In Ref.\cite{KKN}, we have calculated the asymmetries normalized by
the partonic asymmetries, 
$\widetilde{A}_{LL}=-A_{LL}$, 
$\widetilde{A}_{TT}=-A_{TT}/a_{TT}$, 
$\widetilde{A}_{LT}=-A_{LT}/a_{LT}$,
at the center of mass energy $\sqrt{s}=50$ and 200 GeV and 
the virtuality of the photon $Q^2=8^2$ and $10^2$ (GeV)$^2$, which are
within or close to the planned RHIC and HERA-$\vec{N}$ kinematics.
(50 GeV $<$ $\sqrt{s}$ $<$ 500 GeV for RHIC, and 
$\sqrt{s}=40$ GeV for HERA-$\vec{N}$.)
Although the GRSV and GS distributions give quite
different result, the following common features are observed:
\begin{enumerate}
\item[$\bullet$]
$\widetilde{A}_{LT}$ is approximately 5 to 10 times smaller than 
$\widetilde{A}_{LL}$ and $\widetilde{A}_{TT}$.
This is largely due to the factor $x_1$ and $x_2$ in (\ref{ALT}), 
either of which takes quite small values 
in the whole kimematic range considered ($-0.5 < x_F=x_1-x_2<0.5$).
\item[$\bullet$]
The contribution from $\widetilde{h}_L$ and $\widetilde{g}_T$
is almost negligible compared with the Wandzura-Wilczek part,
which is due to the peculiar $Q^2$ evolution shown in Fig.1.
Larger bag scale $\mu^2_{bag}$ would not make 
the former contribution appreciably larger.
\end{enumerate}
From these results, extraction of the effect of the quark-gluon correlation
in $\widetilde{h}_L$ and $\widetilde{g}_T$
would be very challenging in the future polarized Drell-Yan experiment.

Next we discuss briefly the semi-inclusive production of a 
polarized spin-1/2
baryon (mass $M_B$)
in the DIS of an unpolarized electron from the polarized nucleon 
(mass $M$).\cite{KKT}
In this process one can measure the following combination 
of the quark distribution and the fragmentation functions: 
\beq
G_1(x,z)\hskip-0.3cm&=&\hskip-0.3cm
e_a^2 g_1^a(x)\hat{g}_1^a(z),\nonumber\\
G_T(x,z)\hskip-0.3cm&=&\hskip-0.3cm 
e_a^2 \left(g_T^a(x)\hat{g}_1^a(z)
+{M_B h_1^a(x) \hat{h}_L^a(z)\over M x z}\right),\nonumber\\
H_1(x,z)\hskip-0.3cm&=&\hskip-0.3cm 
e_a^2 h_1^a(x)\hat{h}_1^a(z),\nonumber\\
H_L(x,z)\hskip-0.3cm&=&\hskip-0.3cm 
e_a^2 \left(h_L^a(x)\hat{h}_1^a(z)
+{M_B g_1^a(x) \hat{g}_T^a(z)\over M x z}\right),\nonumber
\eeq
where the summation over quark flavors is implied.
Consider the cross section in the target rest frame.
We take the cordinate system in which the lepton beam
defines the $z$-axis and the $x$-$z$ plane contains the nucleon polarization
vector which has a polar angle $\alpha$.  The scattered lepton
has polar and azimuthal angle $(\theta,\phi)$.  
The cross section for the production of the longitudinally polarized
spin-1/2 baryon is
\beq
\hskip-0.5cm& &\hskip-0.5cm{d\Delta\sigma\over dxdydzd\phi}\nonumber\\
\hskip-0.3cm&=&\hskip-0.3cm{\alpha_{em}^2\over Q^2}
\left[ 
-\cos\alpha {1+(1-y)^2\over y}G_1(x,z)\right.\nonumber\\
\hskip-0.3cm&+&\hskip-0.3cm\left.
\sin\alpha\cos\phi\sqrt{(\kappa-1)(1-y)}\right.\nonumber\\
\hskip-0.3cm&\times&\hskip-0.4cm\left.
\left\{
{1+(1-y)^2\over y}G_1(x,z)-{2(2-y)\over y}G_T(x,z)
\right\}
\right].\nonumber
\eeq
Here $x=Q^2/2P\cdot q$, $y=(E-E')/E$, $\kappa=1+(4M^2x^2/Q^2)$,
$z=P\cdot P_B/P\cdot q$, where
$P=(M,0)$, $P_B$, $q$ are the 4-momenta of the initial nucleon,
the produced baryon, and the virtual photon, and 
$E$ and $E'$ are the energies of the initial and the scattered leptons, 
respectively.  Note the appearence of the factor $\sqrt{\kappa-1}\sim O(1/Q)$
in the second term which characterizes twist-3 effect. 
Even at $\alpha=\pi/2$, 
the nucleon's polarization vector is not completely orthogonal to
the virtual photon's momentum and 
the cross section receives the contribution from $G_1(x,z)$, which can be
extracted by setting $\alpha=0$. 

Similarly, the cross section 
for the production of the transversely polarized
baryon is
\beq
\hskip-0.5cm& &\hskip-0.5cm{d\Delta\sigma\over dxdydzd\phi}\nonumber\\
\hskip-0.3cm&=&\hskip-0.3cm{\alpha_{em}^2\over Q^2}
\left[ 
-2\sin\alpha\cos(\phi-\phi') {1-y\over y}H_1(x,z)\right.\nonumber\\
\hskip-0.3cm&+&\hskip-0.3cm\left.
\cos\alpha\cos\phi'\sqrt{(\kappa-1)(1-y)}\right.\nonumber\\
\hskip-0.3cm&\times&\hskip-0.4cm\left.
\left\{
-{2(1-y)\over y}H_1(x,z)+{2(2-y)\over y}H_L(x,z)
\right\}
\right],\nonumber
\eeq
where $\phi'$ is the azimuthal angle between
the polarization vector of the produced baryon $\vec{S}_B$ and
the momentum of the outgoing lepton $\vec{k}'$. 

These semi-inclusive processes and the polarized Drell-Yan
are complementary to each other 
to extract twist-3 distribution functions.

\vspace{0.5cm}
\noindent 
{\bf Acknowledgement.}
I thank Y. Kanazawa, N. Nishiyama and K. Tanaka for the collaboration
on the topics discussed here.  I'm also greatful to RIKEN-BNL center
for the financial support.

\end{document}